\documentclass{iopjournal}

\usepackage[numbers]{natbib}
\usepackage{amssymb}
\usepackage{lipsum}
\usepackage{amsmath}
\usepackage{multirow}
\usepackage{tabularx}
\usepackage{comment}
\usepackage{subcaption}

\begin{document}

\articletype{Paper} %	 e.g. Paper, Letter, Topical Review...

\title{Validating a BDT-based Electron-Positron Identification Algorithm at CLAS12 with Experimental Data}

\author{Mariana Tenorio~Pita$^1$\orcid{0009-0003-7645-1289},Pierre Chatagnon$^{2,*}$\orcid{0000-0002-4705-9582} and Richard Tyson$^3$\orcid{0000-0002-0635-4198}}

\affil{$^1$Old Dominion University, Norfolk, Virginia 23529, USA}

\affil{$^2$Universit\'{e} Paris-Saclay, CEA, IRFU, 91191 Gif-sur-Yvette, France}

\affil{$^3$SUPA, School of Physics and Astronomy, University of Glasgow, Glasgow G12 8QQ, United Kingdom}

\email{marianat@jlab.org, pierre.chatagnon@cea.fr, tyson@jlab.org}


\begin{abstract}
This article presents a machine-learning, particle-identification algorithm for electrons and positrons in the CLAS12 experiment. The main objective was to minimize charged-pion contamination for both experimental and simulated data. We developed, evaluated and validated two BDT models with different input features, both trained on simulated samples, and conducted rigorous validation using both simulated and experimental data to ensure reliability. Our results show the effectiveness of the applied models in mitigating background contamination. By retaining more than 90\% of leptons in the simulated samples, the charged-pion background is largely reduced.  Performance was evaluated on experimental data, providing a framework to validate such approaches at CLAS12 and future electron-beam facilities.
\end{abstract}

\section{Introduction}
The CLAS12 detector~\cite{BURKERT2020163419} is located in the experimental Hall~B at the Thomas Jefferson National Accelerator Facility~(JLab). It is a large-acceptance fixed-target spectrometer constructed as part of the 12-GeV upgrade of the CEBAF accelerator~\cite{Adderley:2024czm}. A continuous electron beam with energy close to 11~GeV is delivered to CLAS12 by the CEBAF accelerator and scattered on a large variety of target systems: liquid di-hydrogen, liquid-deuterium and solid targets. The design of CLAS12, shown in Fig.~\ref{fig:CLAS12}, provides a broad kinematic coverage and excellent particle identification, which makes it particularly well suited for measurements of inclusive and exclusive electroproduction reactions. In this article, we present a BDT-based approach for the identification of electrons and positrons with CLAS12.

\section{The CLAS12 experiment}
\begin{figure}[hbtp]
    \centering
     \begin{subfigure}{0.49\linewidth}
\centering
   \includegraphics[width=\linewidth]{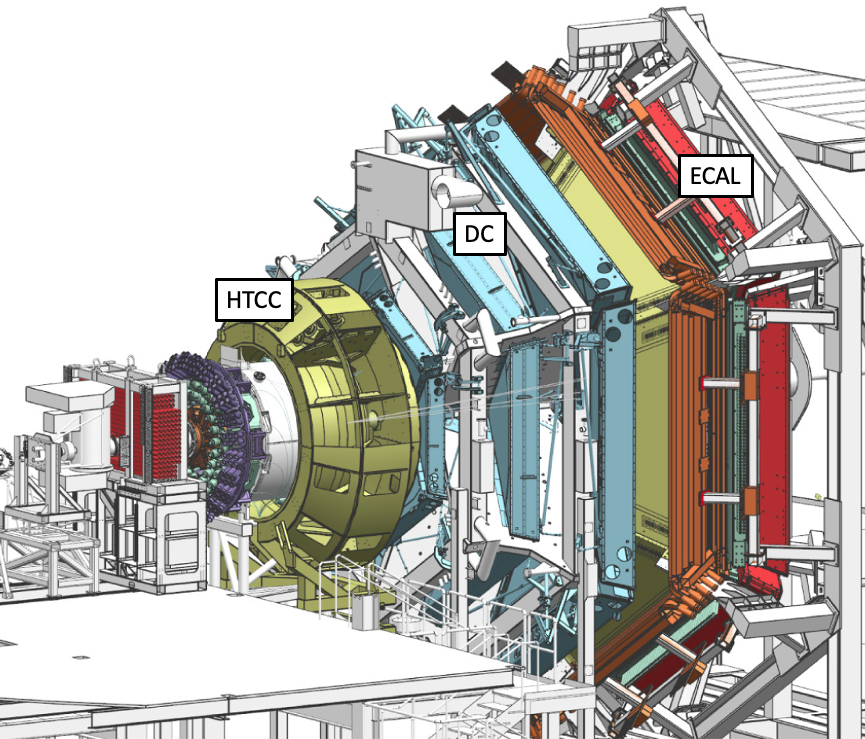}
      \caption{}
\label{fig:CLAS12}
\end{subfigure}
\begin{subfigure}{0.49\linewidth}
\centering
   \includegraphics[width=\linewidth]{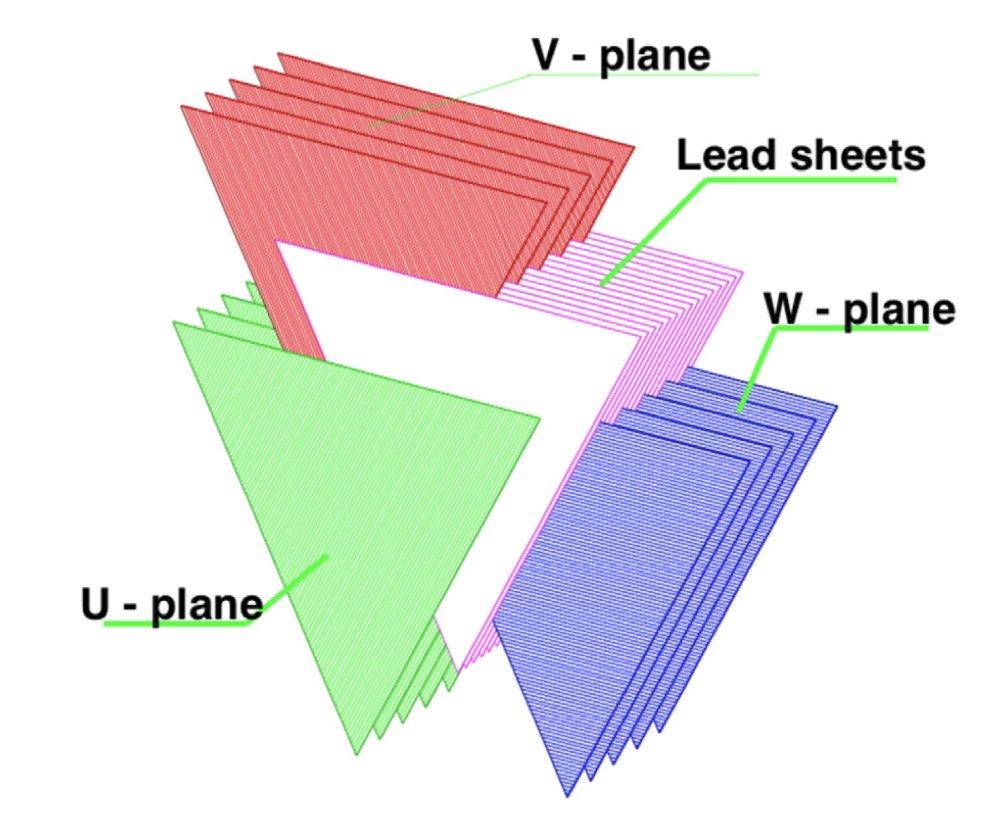}
      \caption{}
\label{fig:calo_layout}
\end{subfigure}
     \caption{Fig.~\ref{fig:CLAS12}: Diagram of the CLAS12 detector. The HTCC~(yellow), DC~(blue), and ECAL~(red) are involved in the electron and positron identifications and are clearly highlighted. Figure adapted from~\cite{BURKERT2020163419}. Fig.~\ref{fig:calo_layout}: Schematic representation of the CLAS12 calorimeter, with the three views (U,V and W) of each layer. Figure taken from~\cite{Asryan:2020iqj}.}
    \label{fig::detector}
\end{figure}

The angular acceptance of CLAS12 is divided into two main regions. On the one hand, the Central Detector (CD) surrounds the target system and is used to detect heavy particle recoiling at large polar angles. On the other hand, the Forward Detector~(FD) is used to identify particle with polar angles from $5^\circ$ to $35^\circ$, and in particular electrons and positrons. It is built around a six-coil superconducting torus magnet~\cite{Fair:2020yfx}. In the FD, charged-particle tracks are reconstructed by the Drift Chamber~(DC) system~\cite{MESTAYER2020163518}, which provides a relative momentum resolution better than 1\% for most data taking conditions. Lepton identification in the FD is achieved using the High Threshold Cherenkov Counter~(HTCC)~\cite{Sharabian:2020whm} in combination with the forward Electromagnetic Calorimeter~(ECAL)~\cite{Asryan:2020iqj}. The HTCC is a $\rm CO_2$-based Cherenkov Counter and consists of 60 mirror sections readout by 48 photo-multiplier tubes~(PMT). It is designed to distinguish leptons from charged pions for momenta below approximately 4.9~$\rm GeV/c$ with a rejection factor of around 500. The ECAL is a lead-scintillator electromagnetic sampling calorimeter, segmented longitudinally into three sub-systems: the Pre-shower Calorimeter~(PCAL), the Inner Electromagnetic Calorimeter~(ECin), and the Outer Electromagnetic Calorimeter~(ECout). The ECAL uses a triangular hodoscope layout with scintillator layers having alternating stereo readout views named U,V and W, as shown in Fig.~\ref{fig:calo_layout}. The PCAL has 68 strips in U and 62 in V and W. Both the ECin and the ECout have 36 strips in all three views.

\section{Electron and positron identification at CLAS12}
\label{sec:leptonID}

The identification of electrons is critical to the CLAS12 physics program, as most measurements require at least the detection of the scattered electron. Within the CLAS12 event reconstruction software~\cite{ZIEGLER2020163472}, the Event Builder~(EB) collects and organizes the responses from the different CLAS12 subsystems. The EB also performs the standard particle identification~(PID) of CLAS12. In the case of electrons and positrons, PID relies on three main subsystems: the DCs, the HTCC and the ECAL. The tracks reconstructed in the DCs are associated with clusters in the ECAL and the HTCC. The ECAL clusters correspond to the localized energy deposited by the particle, and are reconstructed from hits, which are defined by a strip with energy above a certain threshold. Intermediate quantities called peaks are formed from adjacent hits in a plane, with clusters finally identified at the intersection of a peak in each U,V and W views. The signal in the HTCC corresponds to the Cherenkov light emitted by the particle crossing the gas it contains. HTCC clusters are formed by up to four PMT signals that collect Cherenkov light from adjacent mirrors. 

More precisely, a track is identified as an electron~(resp. positron) if the following conditions are fulfilled:
\begin{itemize}
    \item{a track with a curvature corresponding to a negatively charged particle (resp. positively) in the CLAS12 torus magnetic field,}
    \item{a minimum of two photoelectrons in the HTCC,}
    \item{a minimum energy deposition in the PCAL of $60$ MeV,}
    \item{a total sampling fraction (SF) close to the expected value of 0.25, where SF is defined as
    \begin{equation}
        {\rm SF}=\frac{E_{\rm ECAL}}{P},
    \end{equation}
    with $E_{\rm ECAL}$ the total energy deposited in the ECAL measured in GeV, and $P$ the measured momentum of the particle, in $\rm GeV/c$. A parameterization based on simulated data provides both the mean and standard deviation of the SF in bins of momentum. The measured SF is then required to be within five standard deviations of the parametrized mean for a given momentum.
    }
\end{itemize}

For momenta larger than the HTCC Cherenkov threshold, both leptons and charged pions produce a light signal in the HTCC. Therefore, in this kinematic regime, lepton identification relies exclusively on the calorimeter, and the contamination from charged pion becomes larger. The BDT-based approach presented in the following aims at reducing this contamination, especially in the case of positron identification.

\begin{figure}[hbtp]
    \centering
     \begin{subfigure}{0.51\linewidth}
\centering
   \includegraphics[width=\linewidth]{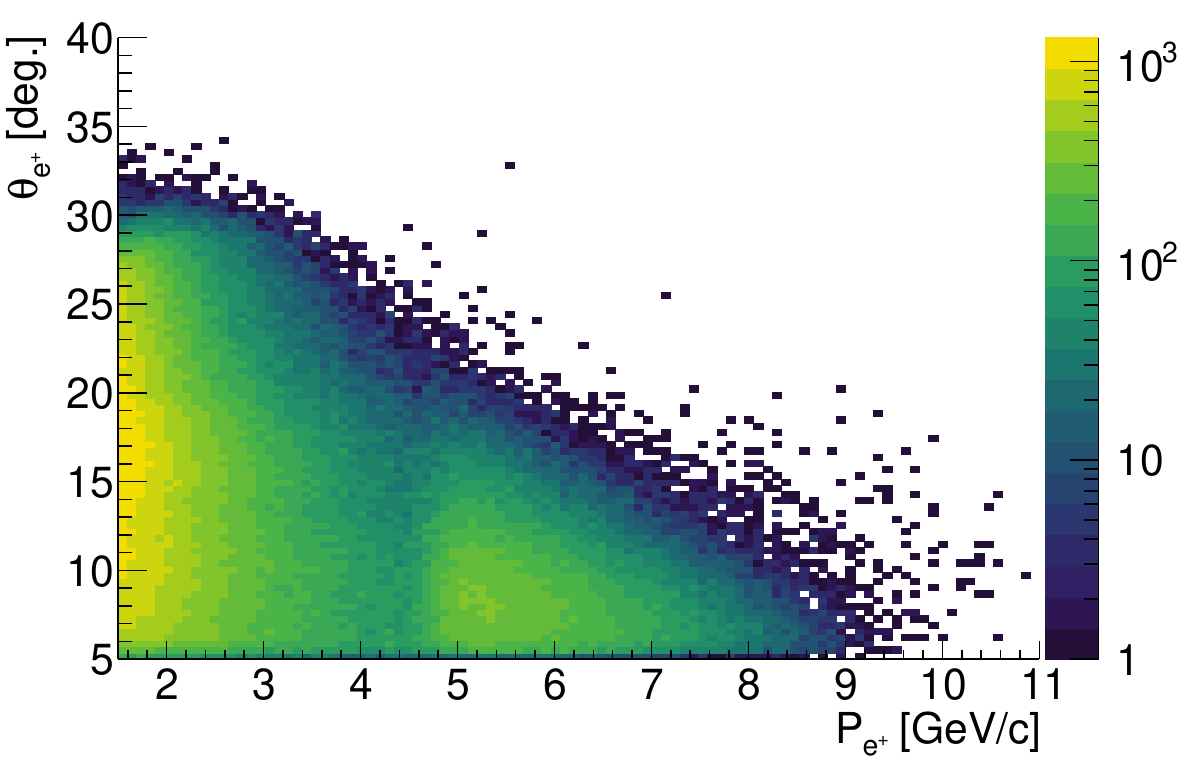}
      \caption{}
\label{fig:evidence1}
\end{subfigure}
\begin{subfigure}{0.48\linewidth}
\centering
   \includegraphics[width=\linewidth]{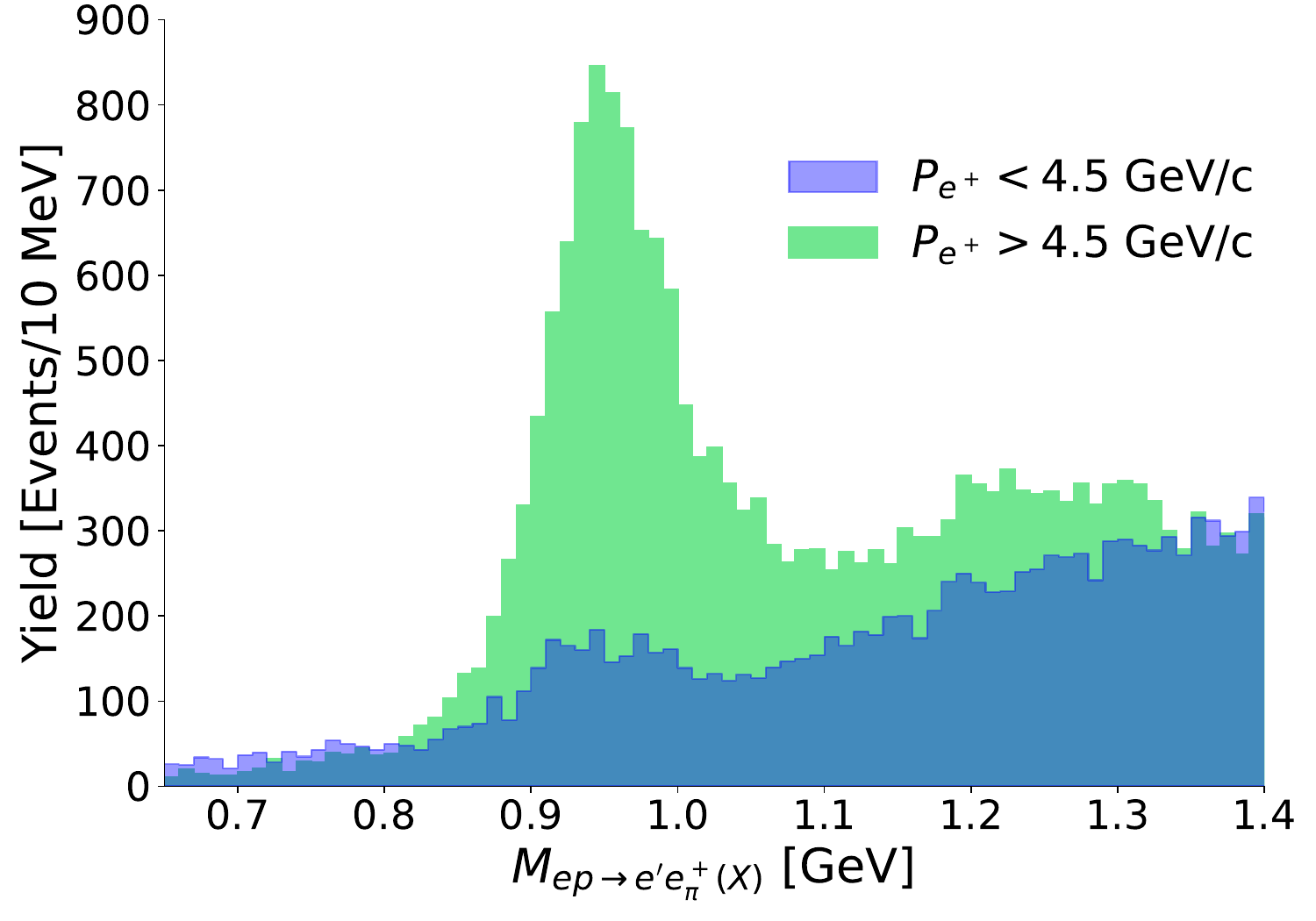}
      \caption{}
\label{fig:evidence2}
\end{subfigure}
     \caption{Fig.~\ref{fig:evidence1}: Positrons polar angle $\theta_{e^+}$ as a function of their momentum $P_{e^+}$. The cluster of events above the HTCC threshold at forward angles indicates a clear contamination from $\pi^+$ falsely identified as positrons. Fig.~\ref{fig:evidence2}: Missing mass distributions of the reaction \mbox{$ep\rightarrow e'e^+_{m_\pi}X$} where the pion mass is assigned to the particle that is identified as a positron by the CLAS12 EB. The green histogram corresponds to positron momenta above \mbox{$4.5$~$\rm GeV/c$}, and a clear peak is visible at the neutron mass due to a large pion contamination in the positron sample. The blue histogram corresponds to positron momenta below \mbox{$4.5$~$\rm GeV/c$}, where the neutron peak is suppressed.}
    \label{fig::evidence}
\end{figure}

To assess the charged-pion contamination in the lepton samples in the CLAS12 datasets, experimental events taken with a liquid-hydrogen target and containing exactly one electron, one positron, and one proton were selected. Figure \ref{fig:evidence1} shows the positron polar angle as a function of its momentum. One can clearly observe a cluster of events above the HTCC threshold at forward angles, indicating contamination from charged pions.

More evidence can be found in the analysis of the exclusive process $ep\rightarrow e^\prime \pi^+ (n)$. In this case, the electron momentum was required to be below $4.5$~$\rm GeV/c$ in coincidence with a positively charged particle identified by the CLAS12 EB as a positron with momentum above $4.5$~$\rm GeV/c$. The latter particle was assigned the mass of charged pions and the missing mass of the reaction \mbox{$ep\rightarrow e'e^+_{m_\pi}X$} was calculated, where $e^+_{m_\pi}$ denotes a positron candidate that was assigned the mass of the charged pion. Figure \ref{fig:evidence2} shows the missing mass $M_{ep\rightarrow e'e^+_{m_\pi}(X)}$ distribution, and a clear peak is visible at the mass of the neutron. This peak can only be explained if the positron candidate is instead a charged pion. The distribution for events in which the identified positron has a momentum below $4.5$~$\rm GeV/c$ is also shown. In this case, the neutron peak is reduced. The momentum threshold below which most leptons are reliably identified is set at 4.5~$\rm GeV/c$. This choice also ensures that the kinematic region near the HTCC threshold is fully accounted for by the algorithm presented later. Finally, this pion contamination was also observed in simulated data. A sample of $\pi^+$ was generated with flat angular and momentum distributions, and passed through the GEANT4-based~\cite{AGOSTINELLI2003250} CLAS12 simulation package~\cite{UNGARO2020163422}. About $10\%$ of positively charged pions with momenta larger than $4.5$~$\rm GeV/c$ were identified as positrons. 

\section{A Machine Learning Approach to Lepton identification}

Machine Learning (ML) methodologies have been extensively used in Particle and Nuclear physics in recent years. In particular, the CLAS collaboration has already developed several ML-based tools, including track identification and denoising~\cite{Gavalian:2022hfa, Thomadakis:2022zue, Thomadakis:2023ebe}, clustering~\cite{Matousek:2025jct}, particle identification~\cite{McEneaney:2023vwp, Matousek:2024vpa}, online reconstruction~\cite{Tyson:2026xuv}, triggering~\cite{Tyson:2023zkx} and unfolding~\cite{Alghamdi:2023emm}. In this work, the goal was to reduce the contamination of charged pions in lepton samples. The confusion matrix for the identification of leptons was defined according to the conventions of Table~\ref{tab:confusionMatrix}. 
\begin{table}[hbtp]
    \centering
        \begin{tabular}{|c|c|c|}
            \cline{2-3}
            \multicolumn{1}{c|}{\rule{0pt}{1.2em}} & True $e^\pm$ & True $\pi^\pm$ \\
            \hline
            Identified $e^\pm$   & True Positive (TP)   & False Positive (FP) \\
            \hline
            Identified $\pi^\pm$ & False Negative (FN)  & True Negative (TN) \\
            \hline
        \end{tabular}
    \caption{Confusion matrix for the lepton identification. True Positive~(TP) events are referred to as signal, while False Positive~(FP) events are referred to as background.}
    \label{tab:confusionMatrix}
\end{table}
The signal consisted of True Positive~(TP) leptons, i.e. genuine leptons that were correctly identified as leptons, while the background consisted of False Positive~(FP) pions, i.e. charged pions misidentified as leptons with the same charge. True Negatives~(TN) consisted of charged pions that were correctly identified as non-leptons, while False Negatives (FN) consisted of true leptons that were incorrectly identified as non-leptons. The performance of lepton identification was quantified using the following metrics
\begin{itemize}
    \item True Positive Rate: $\rm TPR=\frac{TP}{TP+FN}$,
    \item False Positive Rate: $\rm FPR=\frac{FP}{FP+TN}$.
\end{itemize}

\begin{figure*}[hbtp]
    \centering
     \begin{subfigure}{\linewidth}
\centering
    \includegraphics[width=\linewidth]{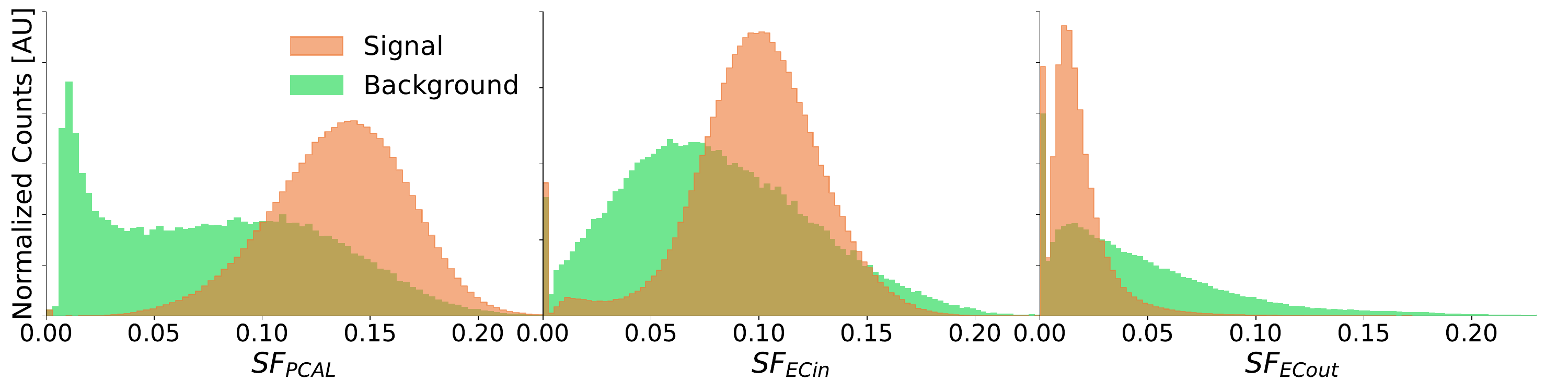}
    \caption{}
\label{subfig:SF_PCAL}
\end{subfigure}
\\
\begin{subfigure}{\linewidth}
\centering
    \includegraphics[width=\linewidth]{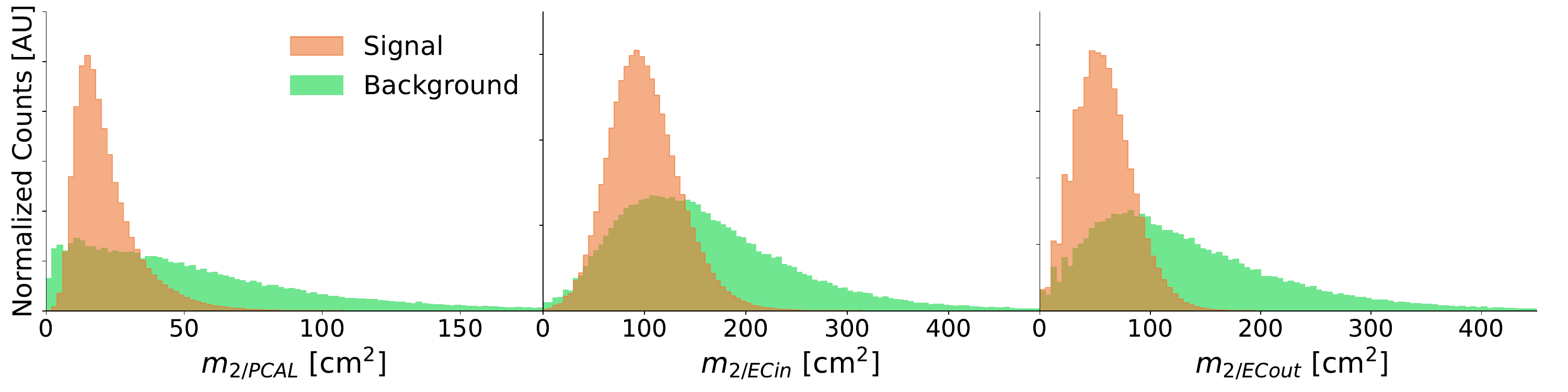}
    \caption{}
\label{subfig:M2_PCAL}
\end{subfigure}

    \caption{Partial SFs (Fig.~\ref{subfig:SF_PCAL}) and transverse size of the EM showers (Fig.~\ref{subfig:M2_PCAL}) of signal (orange) and background (green) obtained from simulated data used to train the lepton identification algorithm. Leptons tend to deposit more energy than mis-identified pions in the PCAL and ECin. The transverse sizes of the EM showers also tend to be smaller for leptons.}
    \label{fig:9BDTinput}
\end{figure*}

We developed a multivariate approach based exclusively on ECAL information, exploiting the different energy deposition distributions of leptons and pions in the ECAL. Electrons and positrons tend to deposit more energy than pions in the PCAL and ECin, and the transverse size of the lepton electromagnetic (EM) shower is typically smaller than that of pions, as shown in Fig.~\ref{fig:9BDTinput}. The CLAS12 ECAL is radially segmented with three independent layers and so the partial SF and the transverse size of the EM showers, $m_2$, in each of the ECAL layers were used as input features.
The transverse size or $2^{nd}$-moments of the EM showers are defined as
\begin{equation}
    m_2=\frac{m_{2U}+m_{2V}+m_{2W}}{3},
\end{equation}
where $m_{2U/V/W}$ are the second-moments of the shower on each stereo readout view of the calorimeter, in cm$^2$, defined as
\begin{equation}
    m_2=\frac{\sum_{\rm hits}(x-D)^2\ln(E)}{\sum_{\rm hits}\ln(E)},
\end{equation}
where $x$ is the position of the hit in the considered view, $E$ is the associated energy of the hit in MeV, and $D$ is the mean position of the shower. All sums are running over hits in a cluster.

The distributions of each input feature obtained from simulated data are shown in Figure \ref{fig:9BDTinput} for both signal and background. One can see that the partial sampling fractions and the transverse size of the EM showers provide a qualitative separation between signal and background. To further exploit these features, Boosted Decision Tree~(BDT) algorithms were developed. Nine input features were used: the partial sampling fraction and the transverse size of the EM showers, as well as the momenta and the polar and azimuthal angles of the particles. These were compared to models using only the sampling fractions and the transverse sizes of the EM showers in order to understand any possible kinematic bias introduced by using momenta and angles. In the following, the models using only six features are referred to as BDT-6, and the full models using the nine features as BDT-9.

The approach described in this article was developed for data taken by the CLAS12 experiment in Fall 2018 and Spring 2019 on liquid-hydrogen targets. During the data taking, the experimental setup remained identical with the exception of the electron beam energy, which was 10.6~GeV for Fall 2018 and 10.2~GeV for Spring 2019. In addition, the Fall 2018 data were acquired with two distinct torus magnet configurations.
The Spring 2019 run and the first part of the Fall 2018 run were taken with the torus magnet set to the inbending configuration at its nominal maximum field strength. In this configuration, negatively charged particles were bent toward the beamline. The second part of the Fall 2018 dataset was collected with the opposite torus polarity, denoted outbending configuration. Therefore, for each model, a total of 6 classifiers were trained: for each data taking configuration and for the two lepton species.

\begin{figure}[hbtp]
    \centering
    \includegraphics[width=0.6\linewidth]{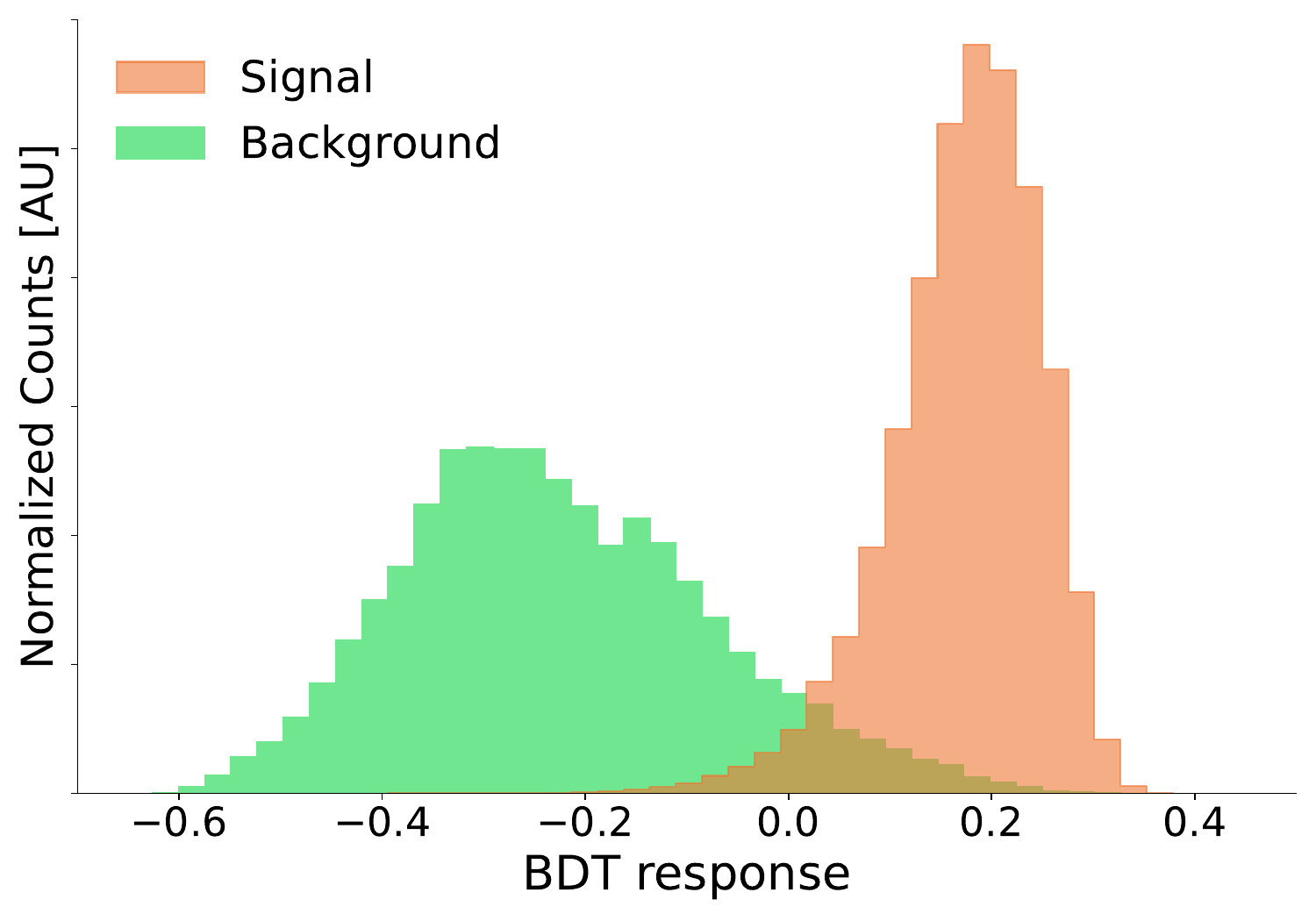}
    \caption{Distribution of the BDT-6 classifier output for signal~(orange) and background~(green) events. The corresponding ROC curve is shown in Fig.~\ref{fig:roc2}.}
    \label{fig::training}
\end{figure}

The ROOT Toolkit for Multivariate Analysis package~\cite{TMVA:2007ngy,Voss:2007jxm} was used to train and evaluate the performance of the classifiers. The BDT models were configured with 850 trees of maximum depth 3, a minimum node size of 2.5\% of the training sample, and 20 cut steps per variable. Boosting was performed with the AdaBoost algorithm using a learning rate of 0.5, bagged boosting was enabled with a sample fraction of 0.5, and the Gini index was used as the node-splitting criterion. The classifiers were trained using simulated samples of electrons, positrons, and charged pions. To fully cover the kinematic region at the HTCC threshold, particles were generated with flat angular distributions and momenta from 4 to 11~$\rm GeV/c$. The samples were then processed using the nominal CLAS12 simulation~\cite{UNGARO2020163422} and reconstruction~\cite{ZIEGLER2020163472} software. Both training samples were built from particles identified as electrons or positrons by the EB, then truth-matched using the generated information. Candidates matching a generated lepton formed the signal sample, and those matching a generated charged pion formed the background. Figure \ref{fig::training} shows the output of the trained BDT-6, which indicates a clear distinction between signal and background, confirming the effectiveness of the method and the separation power provided by the partial sampling fractions and the transverse sizes of the EM showers. 

The optimal cut value on the output of each classifier was obtained by computing the significance 
\begin{equation}
    s=\rm TP/\sqrt{TP+FP},
\end{equation} assuming an equal number of signal and background, and found to be 0. In a physics analysis, the cut value applied to the classifier output can be adjusted depending on the specific requirements of the measurement. Multilayer Perceptrons were also trained using the same input features and training samples. As the training time was longer, and their performances and inference speed were similar to those of the BDTs, the latter were chosen.

\section{Validation and efficiency corrections}
\begin{figure}[hbtp]
    \centering
     \begin{subfigure}{0.49\linewidth}
\centering
    \includegraphics[width=\linewidth]{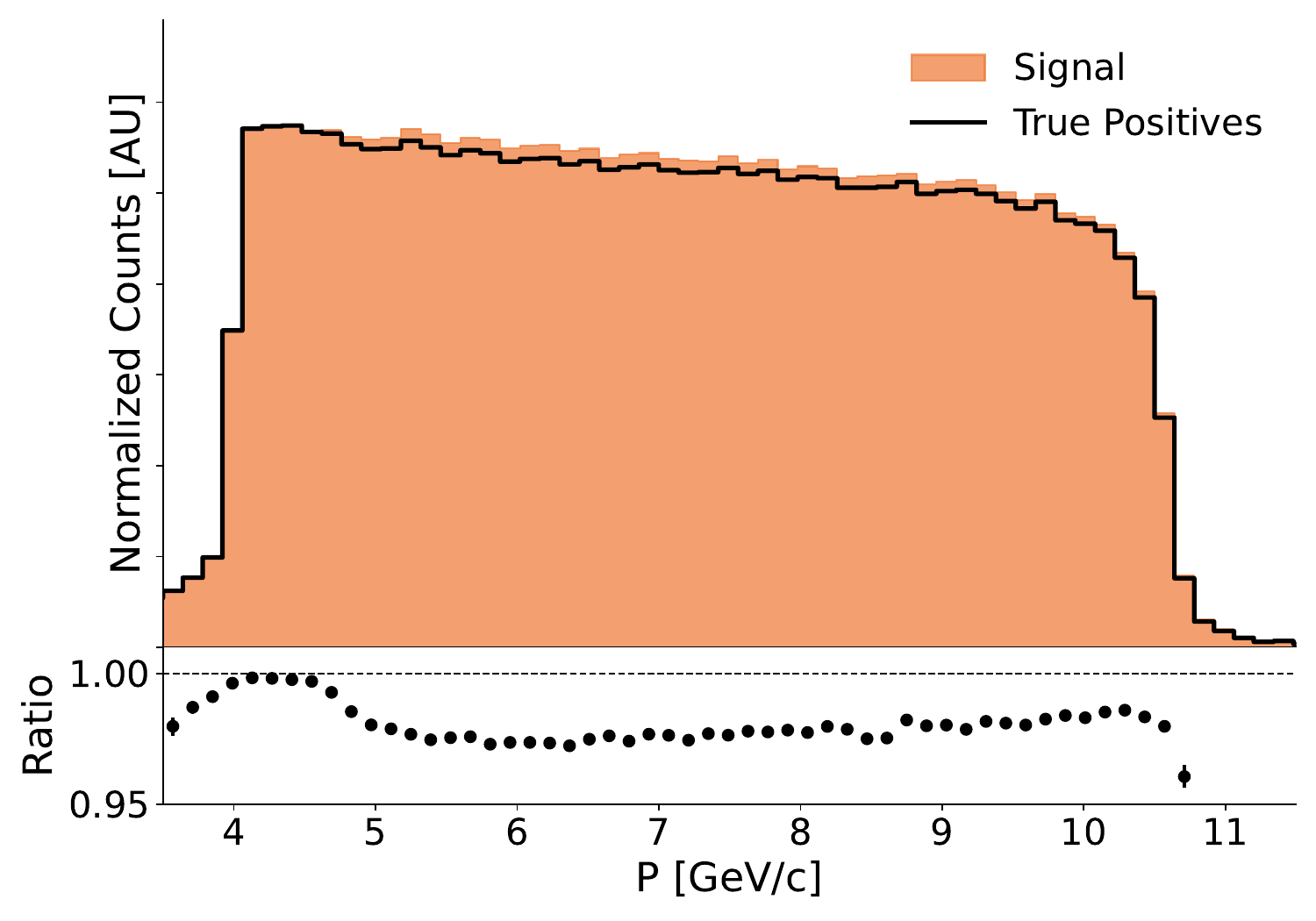}
      \caption{}
\label{subfig:Mom_Sig}
\end{subfigure}
\begin{subfigure}{0.49\linewidth}
\centering
   \includegraphics[width=\linewidth]{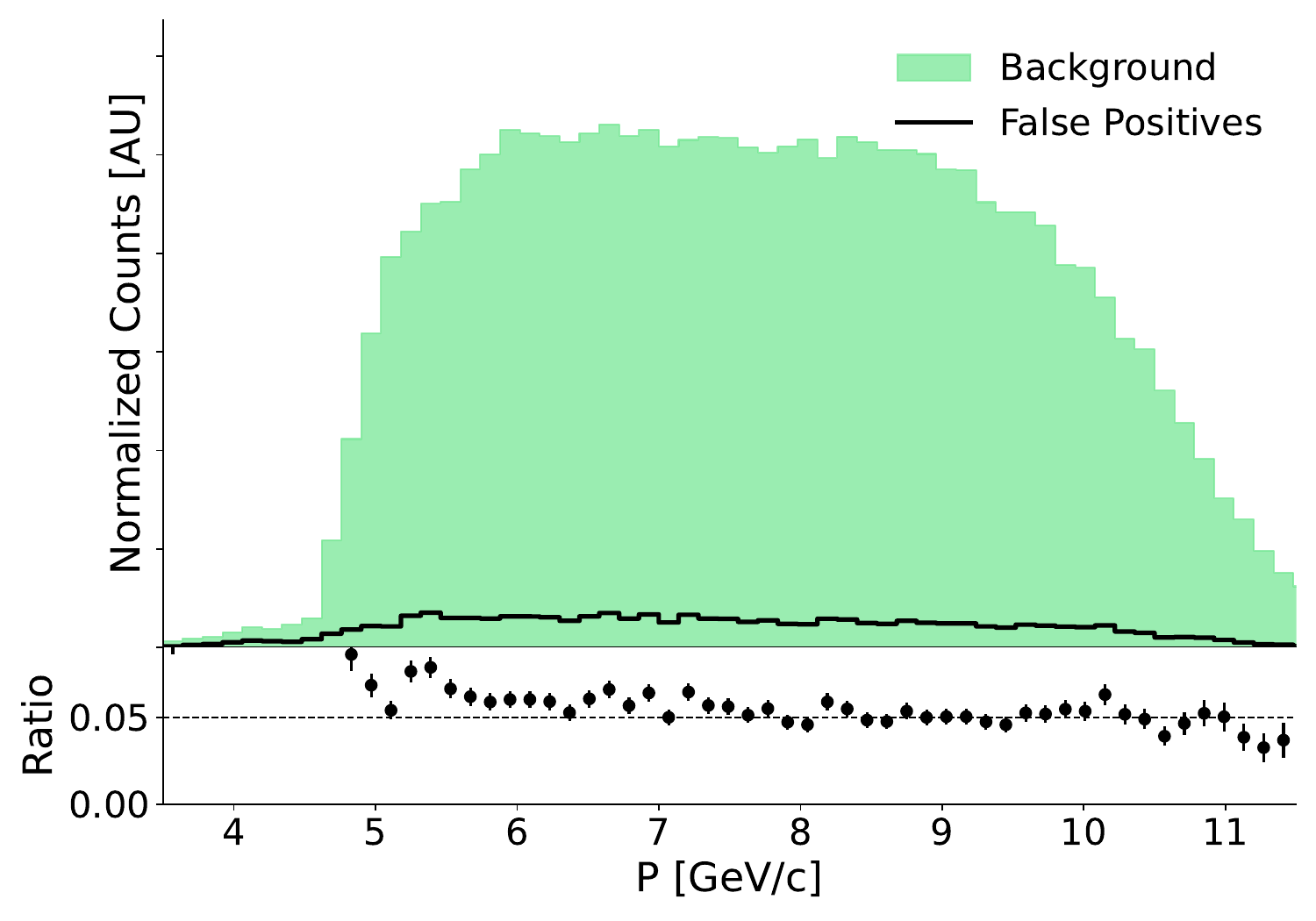}
      \caption{}
\label{subfig:Mom_Bg}
\end{subfigure}
    \caption{Distribution of signal (Fig.~\ref{subfig:Mom_Sig}) and background~(Fig.~\ref{subfig:Mom_Bg}) events as a function of momentum for the initial samples (filled histograms), and after applying the best-selection cut on the BDT-9 classifier (line histograms). The ratio of the number of events after the selection over the number of events in the initial sample is shown with black points under each distribution.}
    \label{fig::val_sim}
\end{figure}

The performance of the classifiers were first evaluated using simulated samples. In particular, ensuring the stability of the classification with respect to the particle kinematics is essential to prevent any bias in subsequent physics analysis. Figures~\ref{subfig:Mom_Sig} and \ref{subfig:Mom_Bg} show the signal and background momentum distributions for electron identification, for the initial samples and after applying a cut at 0 on the output of the BDT-9 classifier, for the Fall 2018 inbending configuration. The ratio of the number of events before and after the cut is shown below each distribution. The integrated TPR for this case is $97.8\%$ and remains constant in the momentum range from 4 to 11~$\rm GeV/c$. The FPR is constantly close to $5\%$. The same validation was performed as a function of polar and azimuthal angles, showing the same stability.

\begin{figure}[hbtp]
    \centering
    \begin{subfigure}{0.49\linewidth}
\centering
  \includegraphics[width=\linewidth]{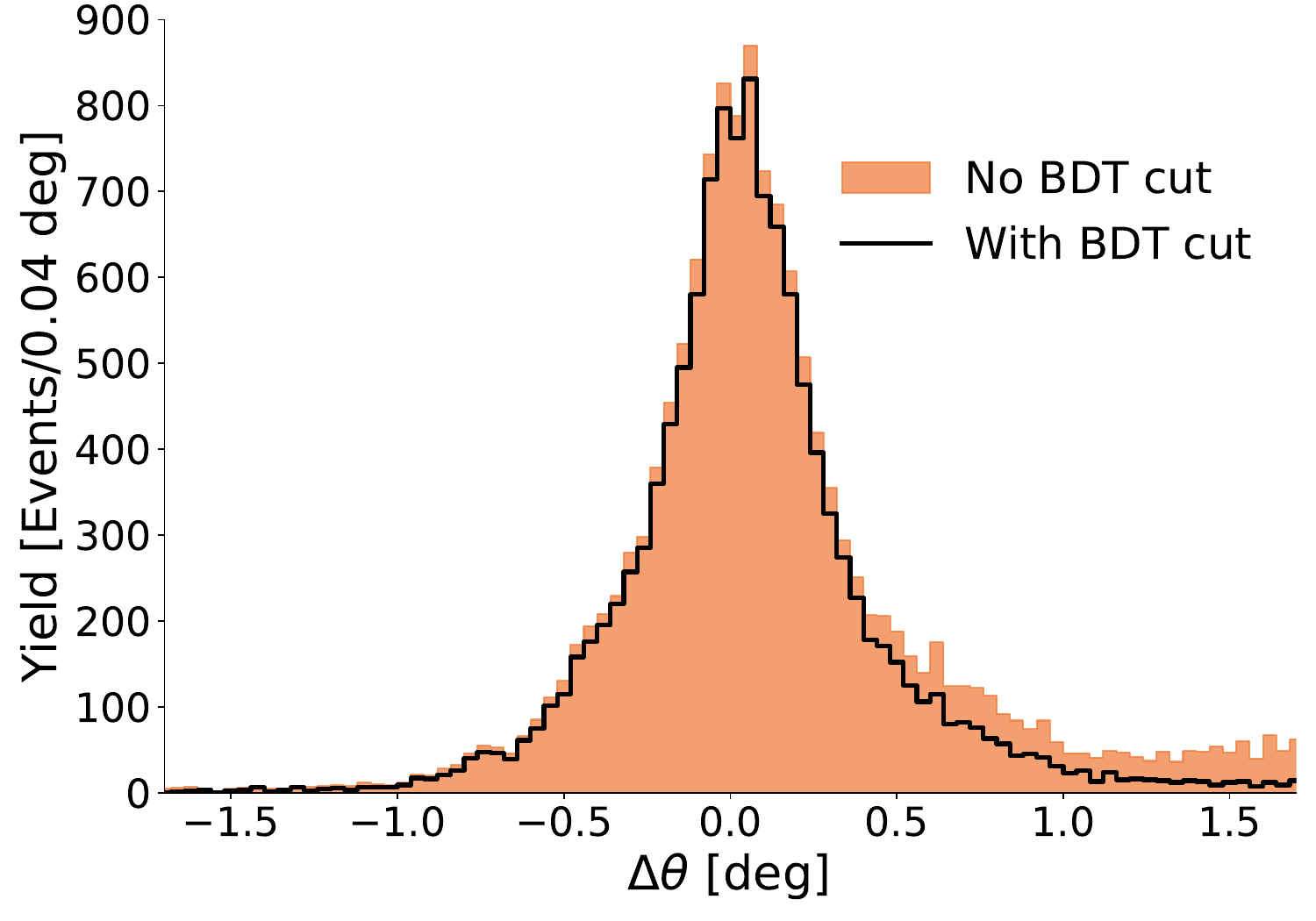}
      \caption{}
\label{subfig:Validation_Signal}
\end{subfigure}
\begin{subfigure}{0.49\linewidth}
\centering
  \includegraphics[width=\linewidth]{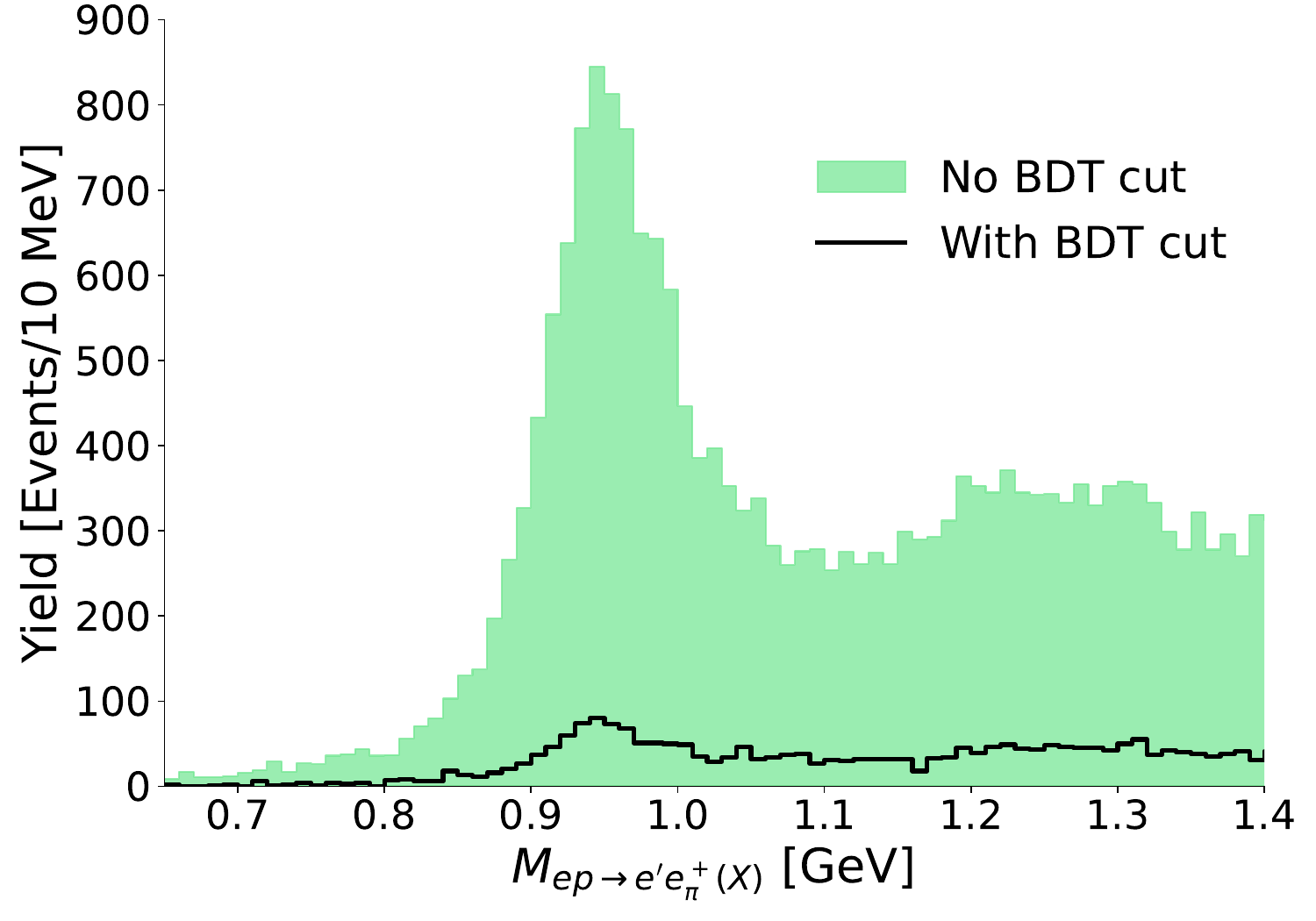}
      \caption{}
\label{subfig:Validation_Bg}
\end{subfigure}
    \caption{Distributions used to validate model performance on experimental data. Figure~\ref{subfig:Validation_Signal} shows the polar angle difference of the electron-photon pair for radiative events. Figure~\ref{subfig:Validation_Bg} shows the missing mass distributions of the reaction \mbox{$ep\rightarrow e'e^+_{m_\pi}X$}. The filled spectra were obtained for the initial samples, without any cut applied. The line histograms were obtained after a selection cut was applied on the output of the BDT-9 classifier. Figure adapted from Ref.~\cite{CLAS:2026lls}.}
    \label{fig::validation}
\end{figure}

The trained models were also extensively validated on experimental CLAS12 data. As electrons and positrons propagate from the CLAS12 target, they can lose energy by radiating photons which can then be detected in the ECAL. In contrast, charged pions do not radiate such photons. Hence, the experimental data signal sample was obtained from these radiative events by selecting the reaction
\begin{equation}
    e^-p\rightarrow e^{-(+)}\gamma X, 
\end{equation}
requiring exactly one electron or positron with momentum greater than $4.5$~$\rm GeV/c$ and one photon. To ensure that the detected photon is emitted by the outgoing lepton, the difference of polar angle of the lepton $\theta_l$ and the photon $\theta_\gamma$ in the laboratory frame was computed as 
\begin{equation}
    \Delta\theta=\theta_\gamma - \theta_l.
\end{equation}
Figure \ref{subfig:Validation_Signal} shows the $\Delta\theta$ distribution for electrons from the selected events. The peak at \mbox{$\Delta\theta=0^\circ$} corresponds to radiative events in which a photon was emitted by an electron. To estimate the TPR of the classifiers in the experimental data, the $\Delta\theta$ distribution was fitted with a Gaussian for the signal peak and a polynomial for the background. The integral of the signal peak for a given cut value $c$, $N^{\rm rad.}_{c}$, was computed and then ${\rm TPR}_{c}$ was estimated as
\begin{equation}
    {\rm TPR}^{\rm data}_{c} = \frac{N^{\rm rad.}_{c}}{N^{\rm rad.}_{\rm tot.}},
\end{equation}
where $N^{\rm rad.}_{\rm tot.}$ is the number of radiative events when no cut was applied. The same procedure was used to estimate the true positron efficiency in the experimental data, although with fewer available statistics.

The selection of a background sample from the experimental data requires a slightly more intricate procedure. For positrons, the reaction described in Section \ref{sec:leptonID} was used, 
\begin{equation}
    ep\rightarrow e'e^+_{m_\pi}X,
\end{equation}
where $e^+_{m_\pi}$ refers to a particle that has been identified as a positron with momentum larger than $4.5$~$\rm GeV/c$ and assigned the mass of the charged pion, and where the missing particle $X$ corresponds to an undetected neutron. As shown in Fig.~\ref{subfig:Validation_Bg}, the visible peak at the neutron mass was fitted and the integral of the neutron peak for a given cut value $c$, $N^{\rm neutron}_{c}$, was computed. The ${\rm FPR}_{c}$ was estimated as
\begin{equation}
    {\rm FPR}^{\rm data}_{c} = \frac{N^{\rm neutron}_{c}}{N^{\rm neutron}_{\rm tot.}},
\end{equation}
where $N^{\rm neutron}_{\rm tot.}$ is the number of neutron events when no cut was applied. As an analog of the reaction above does not exist for electrons, it was instead assumed that electrons ($\pi^-$) and positrons ($\pi^+$) have similar detector responses when varying the polarity of the forward detector toroidal magnet. The outbending dataset was therefore used to estimate the electron FPR in the inbending data set, and vice versa for the positron FPR and the reverse magnetic field polarity. 

\begin{figure*}[hbtp]
    \centering
     \begin{subfigure}{\linewidth}
\centering
    \includegraphics[width=\linewidth]{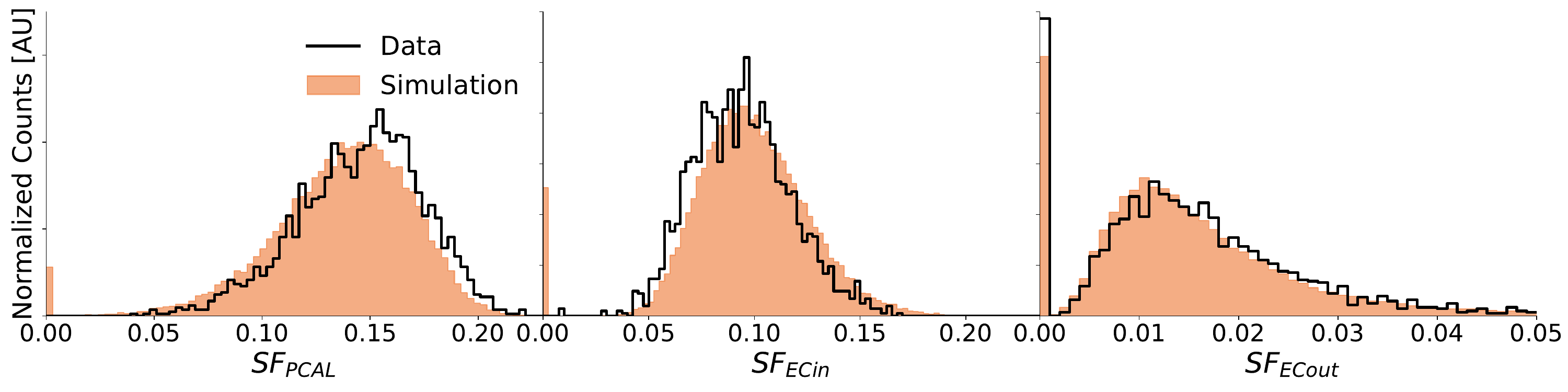}
    \caption{}
\label{subfig:SF_PCAL_MC_DATA}
\end{subfigure}
\\
\begin{subfigure}{\linewidth}
\centering
    \includegraphics[width=\linewidth]{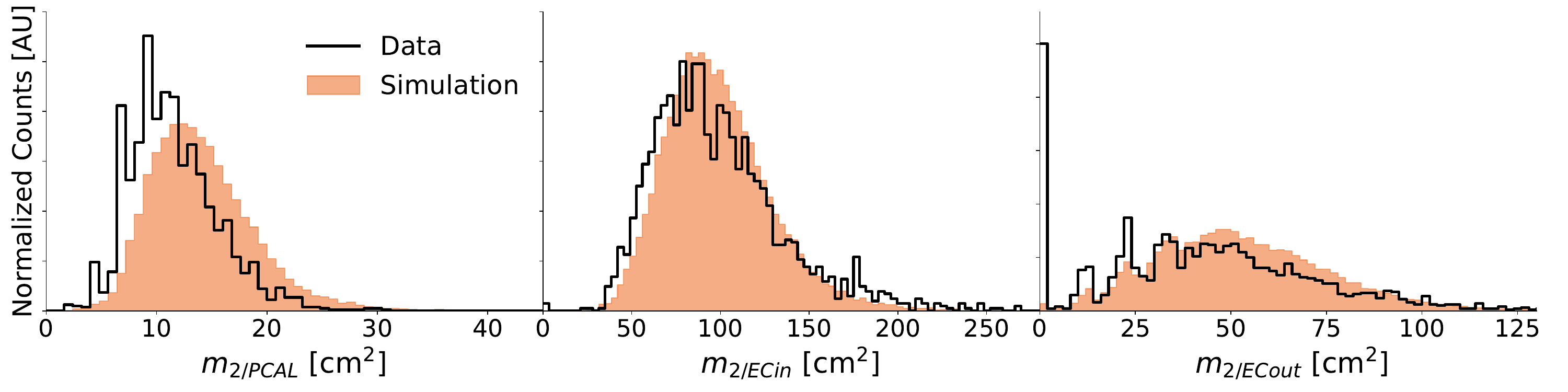}
    \caption{}
\label{subfig:M2_PCAL_MC_DATA}
\end{subfigure}

    \caption{Partial SFs (Fig.~\ref{subfig:SF_PCAL_MC_DATA}) and transverse sizes of the EM showers (Fig.~\ref{subfig:M2_PCAL_MC_DATA}) in each of the ECAL layer for positron signal in simulation (orange filled histograms) and experimental data (line histograms), for the Fall 2018 inbending configuration. Overall, experimental and simulated data spectra are in agreement after positron momentum corrections are applied. }
    \label{fig:var_dist_correction}
\end{figure*}

The calorimeter-related input feature distributions were compared between experimental and simulated data samples to ensure that models trained on simulation maintain reliable performance when applied to experimental data. Figure~\ref{fig:var_dist_correction} shows the partial SFs and the transverse sizes of the EM showers for the simulated signal and for radiative events in data, before and after the best cut on the classifier output was applied. Note that momentum corrections were applied to the experimental data to better match the partial SFs. The spectra from experimental data and simulated samples show a good qualitative agreement, providing confidence in applying the model to experimental data. The differences in efficiencies arising from the remaining mis-modeling were addressed using dedicated efficiency-correction functions, presented later in this section.

The TPR and FPR were evaluated at the best cut value for all models using experimental data, and are summarized in Table~\ref{tab:performances}. The BDT-9 tends to perform better than the BDT-6, while the performance in the outbending configuration tends to degrade for the electron.

\begin{table}[h]
    \parbox{\linewidth}{
        \centering
        \begin{tabular}{|c|c|c|c|c|}
        \cline{2-5}
        \multicolumn{1}{c|}{\rule{0pt}{1.2em}} & Model & Configuration & $\rm TPR$ (\%)  & $\rm FPR$ (\%) \\
        \hline
        \multirow{6}{*}{$e^-$} & \multirow{3}{*}{BDT-6} & Spring 2019 & $96.2 \pm 0.3$ & $8.2 \pm 0.4$  \\
        & & Fall 2018 In. & $96.5 \pm 0.2$ & $11.4 \pm 0.7$   \\
        & & Fall 2018 Out. & $89.3 \pm 2.2$ & $19.0 \pm 1.1$  \\
        \cline{2-5}
        & \multirow{3}{*}{BDT-9} & Spring 2019 & $95.4 \pm 0.3$ & $3.4 \pm 0.3$  \\
        & & Fall 2018 In. & $94.3 \pm 0.2$ & $5.0 \pm 0.5$  \\
        & & Fall 2018 Out. & $81.7 \pm 2.8$ & $11.2 \pm 0.9$  \\
        \hline\hline
        \multirow{6}{*}{$e^+$} & \multirow{3}{*}{BDT-6} & Spring 2019 & $94.2 \pm 0.1$  & $13.8 \pm 0.3$  \\
        & & Fall 2018 Inb. & $94.9 \pm 0.1$ & $12.9 \pm 0.3$  \\
        & & Fall 2018 Out. & $89.7 \pm 0.1$ & $11.0 \pm 0.2$  \\
        \cline{2-5}
        & \multirow{3}{*}{BDT-9} & Spring 2019 & $91.6 \pm 0.1$ & $7.8 \pm 0.3$  \\
        & & Fall 2018 Inb. & $92.2 \pm 0.1$ & $7.1 \pm 0.3$ \\
        & & Fall 2018 Out. & $88.0 \pm 0.1$ & $5.8 \pm 0.2$  \\
        \hline
    \end{tabular}
    \caption{TPR and FPR at the best cut value, estimated using experimental CLAS12 data. The values are given for electrons and positrons, for each data taking period, and for both BDT models presented in this work.}
    \label{tab:performances}
    }
\end{table}

\begin{figure}[ht]
    \centering
    {\label{fig::ROCin}\includegraphics[width=0.6\linewidth]{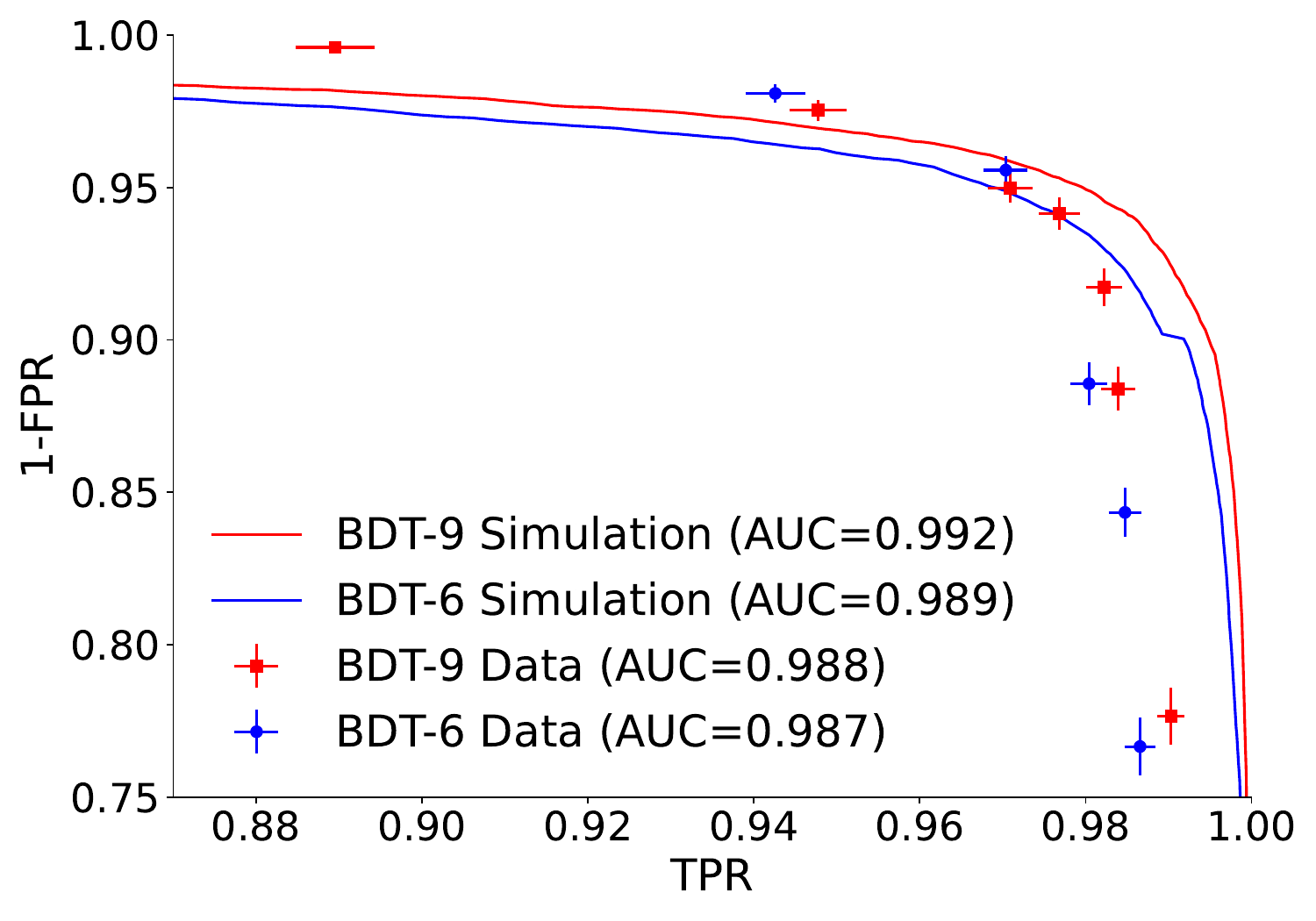}}
    \caption{Comparison of the ROC curves obtained from experimental data (points) and simulation (solid line) for the BDT-6 (blue) and BDT-9 (red) models, and for the Fall 2018 inbending datasets. The associated Area Under the Curve (AUCs) are given for each curve. The TPR from the data are lower than those in simulated data, which was accounted for by dedicated correction functions. }
    \label{fig:roc2}
\end{figure}

Combining the TPR and FPR obtained from experimental data, the ROC curves of the classifiers were calculated and compared with those obtained from simulated data, as shown in Figure \ref{fig:roc2}. The curves derived from experimental data are qualitatively consistent with those from simulated data, with slightly better TPRs in simulated data.

\begin{figure}[hbtp]
    \centering
    \includegraphics[width=0.6\linewidth]{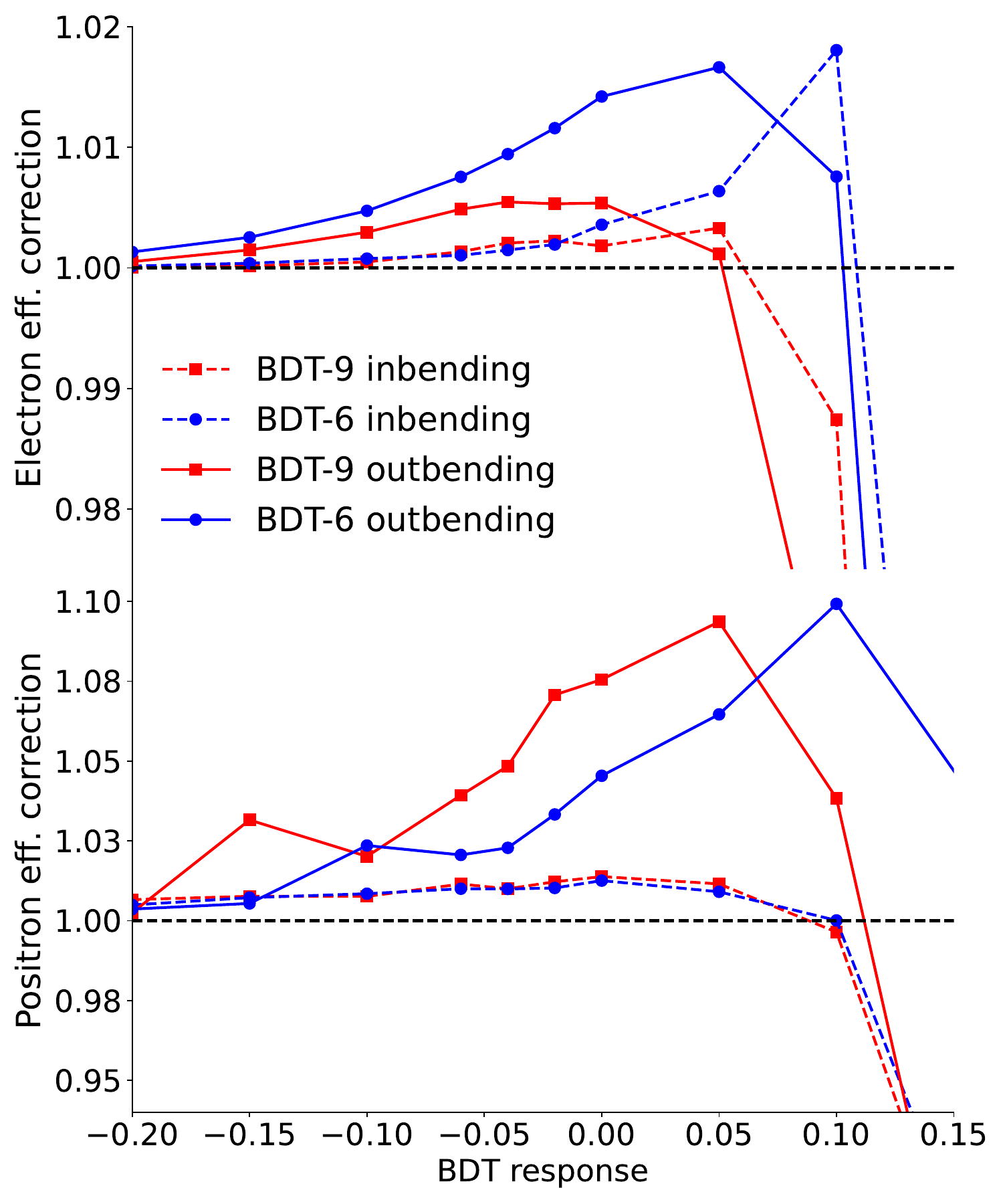}
    \caption{Efficiency correction functions for the Fall 2018 datasets as a function of the classifier output, for electrons (top) and positrons (bottom) for BDT-6 (blue circles) and BDT-9 (red squares) classifiers, and for both inbending (dashed lines) and outbending (solid lines) configurations.}
        \label{fig::ratio_inout}
\end{figure}

To correct for the remaining efficiency mismatches between simulation and experimental data, correction functions $C$ were derived as the ratio of the TPR obtained in simulated data ($\rm TPR^{simu.}_{e^{+/-}}$) to the one obtained in experimental data ($\rm TPR^{data}_{e^{+/-}}$):
\begin{equation}
    C_{e^{+/-}}=\frac{\rm TPR^{simu.}_{e^{+/-}}}{\rm TPR^{data}_{e^{+/-}}}.
\end{equation}
Figure \ref{fig::ratio_inout} shows the correction functions for the different models of the Fall 2018 datasets as a function of the cut applied on the classifier output.  The discrepancy between simulated and experimental data near the best cut value of 0 was found to be less than 2$\%$ for most configurations. The outbending positron case was found to have the largest discrepancy at 8$\%$. These correction functions are essential for CLAS12 analysis involving the detection of leptons, as they were used to correct the efficiency of electron and positron identification in simulated data by multiplying each event by a weight $w$ defined as
\begin{equation}
    w=\left(\frac{1}{C_{e^+}}\right)^{n_{e^+}}\left(\frac{1}{C_{e^-}}\right)^{n_{e^-}},
\end{equation}
where ${n_{e^{+/-}}}$ is the number of positron/electron involved in the reaction under investigation. As detailed in the next section, in the case of cross section calculation, the systematic error associated with the choice of cut value can be assessed by varying the cut value and using the dependence shown in Fig.~\ref{fig::ratio_inout}.

\section{Implementation and usage}
The approach presented in this article has already been used in two CLAS collaboration publications, the first measurement of Timelike Compton Scattering~(TCS)~\cite{CLAS:2021lky} and the measurement of the near-threshold $J/\psi$ photoproduction cross section~\cite{CLAS:2026lls}. Both analyses required the identification of an electron-positron pair in the final state, where the initial electron was not detected:
\begin{equation}
    ep\rightarrow (e')p'e^- e^+.
\end{equation}
The first version of the positron classifier was developed for the TCS analysis where the use of the positron classifier was required to minimize the background in the electron-positron invariant mass region between 1.5 and 3 GeV. The method was further developed and validated for the measurement of near-threshold $J/\psi$ photoproduction. The comparison between simulated and experimental data efficiencies was particularly crucial for the $J/\psi$ cross-section measurement, as the measured yields were corrected by the lepton identification efficiency and any mis-modeling of the efficiency between simulation and experimental data would propagate directly into the extracted cross sections. Ensuring consistency between efficiencies measured in simulation and those observed in experimental data was therefore essential to avoid introducing biases into the final physics results. 

Following the work presented in this article, all classifiers have been integrated into the {\it Iguana} software~\cite{iguana}, which provides a unified analysis framework for CLAS12 data analysis. The approach extends easily to other CLAS12 datasets, including ones taken on deuterium and nuclear targets, by simply training dedicated BDTs for specific datasets.

The validation methodology presented here extends beyond the application of lepton identification tools to individual analyses at CLAS12. Electron identification is a fundamental requirement for experiments employing electron beams, as the scattered electron is present in many physics channels and is often used as the basis for triggering data acquisition. Furthermore, key kinematic quantities, such as the quasi-real photon virtuality, $Q^2$, and the Bjorken scaling variable, $x_B$, are derived from the scattered electron kinematics, making reliable electron reconstruction essential for the physics program of these experiments. The development of advanced electron identification algorithms, such as the approach presented here, therefore has broad potential to enhance experimental capabilities. However, as discussed above, cross-section measurements require a well-established understanding of identification efficiencies, while ML-based triggering approaches, such as those presented in Refs.~\cite{Tyson:2026xuv,Tyson:2023zkx}, require rigorous validation with experimental data to ensure efficient and unbiased data collection. The methods developed here hence provide a validation framework for advanced electron and positron identification techniques, which could be also implemented in future experiments.

\section{Conclusion}
In this article, algorithms based on Boosted Decision Trees were applied to better discriminate between leptons and pions in the CLAS12 experiment at JLab. This approach was shown to reduce the pion contamination in the electron and positron samples by at least an order of magnitude, both in experimental and simulated data. The models developed in this work were trained using simulated datasets. However, to ensure the robustness and reliability of our models, extensive validation procedures were conducted using experimental CLAS12 data. Performance on experimental data was shown to be consistent with that obtained in simulated data, and efficiency correction functions were developed to account for the remaining mismatches, to be later applied to cross section measurements. These tools were developed for the liquid-hydrogen dataset of CLAS12, but can be easily extended to other datasets, as they are now implemented as part of the collaboration-wide {\it Iguana}~\cite{iguana} framework. This work demonstrates the potential of exploiting the full information available from sampling calorimeters for advanced lepton identification, providing a pathway for similar developments in future experiments at JLab, such as SoLID~\cite{JeffersonLabSoLID:2022iod} and $\mu \rm CLAS12$~\cite{Alvarado:2026ggy}, as well as at the future ePIC experiment at the Electron Ion Collider~\cite{AbdulKhalek:2021gbh, Allaire:2023fgp}. In particular, the validation framework presented in this article provides a general approach for establishing the reliability of such machine-learning-based identification methods, ensuring that these techniques can be reliably integrated into future experimental programs.

\ack{The authors thank the CLAS Collaboration for providing the data used in this work. }
\funding{This material is based upon work supported by the U.S. Department of Energy, Office of Science, Office of Nuclear Physics under Contract No. 89243126CSC000213, by the U.S. Department of Energy under Award No. DE-FG02-96ER40960, and upon work supported by the U.K. Science and Technology Facilities Council under grants ST/Z510312/1 and ST/Y000315/1.}
\roles{M. Tenorio: Writing - original draft, Data analysis, Software, Investigation, Validation, Visualization. P. Chatagnon: Writing - editing and review, Conceptualization of this study, Methodology, Software, Visualization. R. Tyson: Writing - editing and review, Methodology.}

\bibliographystyle{unsrt}
\bibliography{ML_Lepton}

\end{document}